# Exploring the robustness of features and enhancement on speech recognition systems in highly-reverberant real environments

*José Novoa* [1], *Juan Pablo Escudero* [1], *Jorge Wuth* [1], *Victor Poblete* [2], *Simon King* [3], *Richard Stern* [4] *and Néstor Becerra Yoma* [1]

[1] Speech Processing and Transmission Laboratory, Electrical Engineering Department, Universidad de Chile, Santiago, Chile.
[2] Institute of Acoustics, Universidad Austral de Chile, Valdivia, Chile.
[3] Centre for Speech Technology Research, University of Edinburgh, Edinburgh, UK.
[4] Department of Electrical and Computer Engineering and Language Technologies Institute, Carnegie Mellon University, Pittsburgh, USA.

nbecerra@ing.uchile.cl

## Abstract

This paper evaluates the robustness of a DNN-HMM-based speech recognition system in highly-reverberant real environments using the HRRE database. The performance of locally-normalized filter bank (LNFB) and Mel filter bank (MelFB) features in combination with Non-negative Matrix Factorization (NMF), Suppression of Slowly-varying components and the Falling edge (SSF) and Weighted Prediction Error (WPE) enhancement methods are discussed and evaluated. Two training conditions were considered: clean and reverberated (Reverb). With Reverb training the use of WPE and LNFB provides WERs that are 3% and 20% lower in average than SSF and NMF, respectively. WPE and MelFB provides WERs that are 11% and 24% lower in average than SSF and NMF, respectively. With clean training, which represents a significant mismatch between testing and training conditions, LNFB features clearly outperform MelFB features. The results show that different types of training, parametrization, and enhancement techniques may work better for a specific combination of speaker-microphone distance and reverberation time. This suggests that there could be some degree of complementarity between systems trained with different enhancement and parametrization methods.

**Index Terms**: speech recognition, enhancement techniques, reverberation, real environments

## 1. Introduction

Distant automatic speech recognition (ASR) represents a major challenge because the effects of noise and reverberation on the speech signal increase as the distance between speaker and microphone increases [1]. Despite recent advances in ASR technology, successful distant speech recognition in real reverberant environments remains an important challenge [2].

Several algorithms have been proposed to address the reverberation problem such as SSF, NMF and WPE. SSF is motivated by the precedence effect. While the precedence effect is clearly helpful in enabling the perceived location of a source in a reverberant environment to remain constant, as it is dominated by the characteristics of the components of the sound which arrive directly from the sound source while suppressing the potential impact of later-arriving reflected components from other directions. It is also believed by some to improve speech intelligibility in reverberant environments as well [3] [4].

While hearing researchers have traditionally modeled precedence using binaural mechanisms [5], it is also possible that onset enhancement at the peripheral level (*e.g.* [6]) may be involved instead. This conjecture motivated the SSF algorithm, which accomplishes this type of onset enhancement and steady-state suppression on a band-by-band basis [7]. Onset enhancement is accomplished by nonlinear extraction of the lower envelope and subtracting it from the ongoing envelope of components of the input after peripheral bandpass filtering, followed by suppression of the steady-state frames after the initial arrival of a wavefront in a particular frequency band. There is a subsequent spectral reshaping module that minimizes differences between the power spectra representing the original and processed speech. While the original description of SSF described two modes of processing, we employ only the processing referred to as SSF Type II in this work. In speech recognition applications, SSF processing is normally applied to both training and testing data.

Non-negative matrix factorization, or NMF, is very different in nature, in that in effect it accomplishes blind deconvolution of the response to a reverberated signal in the frequency domain [8]. It is easy to observe that the presence of reverberation causes a representation like a spectrogram to become blurred or smeared along the time axis, caused by convolution of the response representing clean speech with the sample response of the room acoustics, as represented in the frequency domain. Because phase information is lost in the spectrogram, blind deconvolution cannot be accomplished exactly, but a good approximation can be achieved by exploiting the facts that the matrix representing the sample response in the frequency domain would be non-negative and sparse.

The implementation of NMF that is used [8] differs from previous work in this area in several ways. First, processing is performed using magnitude spectra rather than power spectra. We have found that the distortion introduced by the approximation of non-negativity is reduced through the use of magnitude spectral (rather than power spectral) coefficients. A second innovation is the use of a frequency representation based on Gammatone filtering (which mimics the peripheral frequency analysis of the human auditory system) rather than a conventional linear or log-based frequency distribution. The use of the Gammatone sub-bands provides a natural perceptual weighting to the optimization process which has proved to be helpful, and at the same time it reduces the amount of computation that is required. Typically, the matrix representing the reverberation filter is estimated on a band-by-band basis using an objective function that includes constraints based on both sparsity and non-negativity. Further details may be found in [8].

Weighted Prediction Error algorithm (WPE) [9] [10] is another enhancement technique. This method is based on robust blind deconvolution using long-term linear prediction, with the motive of reducing the effects of late reverberation. This method receives as input a speech signal in the time domain. A complex STFT is performed to compute the coefficients of the linear prediction filters iteratively. Finally, a de-reverberated time waveform is obtained.

It has been found, unsurprisingly, that ASR systems exhibit the best performance when training and testing conditions are matched. Correspondingly, acoustic models produce more errors if test data are different in nature from training data. This degradation can occur, for example, if clean data are used for training but test data are corrupted by noise [11]. In cases of severe mismatches between training and test conditions, allowing the DNNs to see examples of representative variations during training can provide improvements in performance as the DNN can potentially extract useful information from those examples through the layers of nonlinear processing. In this way, the DNN is able to generalize to similar patterns in the testing data [12] enabling the DNN to become less sensitive to changes of the input [11]. For this reason, a typical way to achieve noise robustness of DNN is by using multi-style training. A DNN trained with several noise types and SNR levels can lead to improvements in ASR performance in noisy environments [12].

In addition to additive noise, reverberation is one of the major sources of mismatch between training and testing conditions, and hence also degrades recognition accuracy. An efficient way of mitigating this mismatch is to train models using Multi-condition/Multi-style training data [13]. Multi-style training creates matched training/test environments by adding background noise and/or simulated reverberations to the data used to train the models. This method is effective in compensating for the effects of mismatch [14]. This approach has been reported by multiple research groups at the 2014 REVERB Challenge workshop [15], showing that performing Multi-condition training using a variety of reverberation conditions usually improves the robustness of acoustic models [16]. Similarly, generalized Multi-style training was used in [2], where the network is provided with a characterization of reverberation in which the test data was captured in order to build room-awareness into the model. An NMF-based method was used to estimate a non-negative representation of the clean speech signal and the room impulse response directly from the reverberant speech. This technique also leads to significant improvement when evaluated using the REVERB Challenge corpus.

All of the reverberated speech databases (e.g. REVERB challenge [15], CHiME-2 Challenge [17], and ASpIRE [18]) that have been employed so far attempt to use real environments and, in most cases, also include additive noise. Surprisingly, the impact of reverberation time (RT) and speaker-microphone distance on the performance of ASR technologies has not yet been addressed methodologically and independently of the additive noise. This is partly because there has not been a suitable database for this purpose. The HRRE database [19] is a response to this need by providing speech data recorded in a controlled reverberant environment for several different speaker-microphone distances. By doing so, we cover a wide range of potential applications that include human-robot applications, meeting rooms, smart houses to close-talking microphone scenarios.

An alternative approach to address the reverberation problem is the design or use of robust features. In [20], Damped Oscillator Coefficients (DOC), Normalized Modulation Coefficients, Modulation of Medium Duration Speech Amplitudes, and Gammatone Filter Coefficients features were evaluated on the REVERB 2014 challenge data. In [21], Gammatone filterbank and DOC features were testing under reverberated conditions on the ASpIRE task [18].

A novel set of speech features for robust Speaker Verification (SV) and ASR called Locally-Normalized Cepstral Coefficients (LNCC) was proposed in [22]. LNCC features are inspired by Seneff's Generalized Synchrony Detector (GSD) [23] which performs a local normalization in the frequency domain in each auditory channel, and hence is relatively invariant to changes in the frequency response of the transmission channel. LNCC features are an extremely simple but effective way to instantaneously normalize speech features with respect to frequency. Their effectiveness was demonstrated in a SV task in which LNCC features were more effective in compensating for spectral tilt [22] and more robust to additive noise [24] compared to ordinary MFCC coefficients.

The comparison of different robust features in combination with enhancement techniques in a controlled highly-variable real reverberant environment is not found in the literature. In this paper the robustness of LNFB and MelFB features in combination with NMF, SSF and WPE enhancement methods is discussed and evaluated regarding RT and speaker-microphone distance with clean and reverberated training. The results presented here suggests that there might be some degree of complementarity between systems with different training strategies, enhancement techniques and parametrizations.

## 2. Experiments

### 2.1. Training data

Speech recognition experiments were performed using the Kaldi Speech Recognition Toolkit [25]. The Clean training set from the Aurora-4 database was employed. This set contains 7138 utterances from 83 speakers recorded with a Sennheiser HMD-414 microphone. Additionally, a reverberant training set was developed by our group, referred to as "Reverb."

For Reverb training, simulations were made with the simulation program Room Impulse Response Generator [26], which uses the image method assuming a rectangular room [27]. In order to avoid potential artifacts in training because of potential standing wave patterns that may develop in rectangular rooms, the Reverb training database consists of 5353 utterances that were passed through 5353 different randomly-generated room impulse responses (RIRs). The dimensions of the simulated rooms varied from RIR to RIR with an average of 7.95 m length, 5.68 m width and 4.5 m height, approximating the dimensions of the larger-sized reverberation chamber of the Acoustic Institute. The dimensions for each individual RIR were drawn from uniform distributions over the range of plus or minus 20 percent of the nominal values stated above. A nominal RT was then selected by sampling a random variable over the range of 0.45 to 1.87 s, and the nominal average absorption and reflection coefficients that would provide the selected nominal RT were calculated using the Sabine equation [28]. Six separate reflection coefficients, one for each room surface, were drawn from a uniform distribution between plus and minus 10 percent of the nominal reflection coefficient calculated from the Sabine equation, resulting in a room with a reverberation that was random, but close to the intended nominal value. The distance between speaker and microphone was drawn from a uniform distribution between 0.144 and 2.816 m. The speaker and microphone were placed in random locations at the room, using the distance that was selected for a particular trial, with the constraints that both speaker and microphone are at least 1 m from any wall and between 1 m and 2 m from the floor.

### 2.2. System training

Two types of feature vectors were compared in this paper, the MelFB and LNFB features, in both cases considering a context window of 11 frames, including 5 frames before and 5 frames after the current frame. Each DNN in the DNN-HMM system consists of seven hidden layers and 2048 units per layer. The DNN-HMM systems were trained using alignments from an GMM-HMM recognizer trained with the same data. In turn, the GMM-HMM systems were trained by using MFCC features, linear discriminant analysis (LDA), and maximum likelihood linear transforms (MLLT), according to the tri2b Kaldi Aurora-4 recipe. First, a monophone system was trained; second, the alignments from that system were employed to generate an initial triphone system; and finally, the triphone alignments were employed to train the final triphone system. The number of units of the output DNN layer was equal to the number of Gaussians in the corresponding GMM-HMM system. For decoding we used the standard 5K lexicon and trigram language model.

## 3. Results and discussion

We obtained results for a total of 330 testing utterances for each one of the 20 reverberation conditions (four RTs and five microphone-speaker distances) available in the HRRE database [19]: RTs equal to 0.47s, 0.84s, 1.27s, and 1.77s; and, microphone-speaker distances equal to 0.16m, 0.32m, 0.64m, 1.28m, and 2.56m. Two types of feature extraction procedures (MelFB and LNFB), two sets of training data (Clean and Reverb) and four types of environmental compensation (none, NMF, SSF, and WPE) were combined.

Table 1 describes the WERs obtained for each speaker-microphone distance averaged across the four RTs that were available in our reverberation chamber. The lowest WER for each column is highlighted in bold in Table 1. As can be seen in Table 1, the best results are observed for Reverb training with MelFB combined with WPE in most cases. The best MelFB features perform better than the best LNFB features (in conjunction with Reverb training) averaged over all RTs. Compared with the baseline system with MelFB and Clean training condition, the optimal reductions in Table 1 are higher than 70% with all the speaker-microphone distances.

### 3.1. Training procedure

According to what has been mentioned about multi-style training, the best results are achieved with Reverb training in most test conditions. However, as can be seen in Fig. 1, Clean training in combination with WPE achieves better performance than Reverb training in four of the twenty conditions: RT equal to 0.84s and 1.27s at 2.56 m using LNFB; and, with RT equal 0.47s in the shortest distances (*i.e.* 0.16 and 0.32 m) using MelFB.

Table 1: *WERs averaged across all RTs values using MelFB and LNFB for different training conditions and pre-processing techniques.*

| Training | | Feature | Speaker-microphone distance [m] | | | | |
|---|---|---|---|---|---|---|---|
| | | | 0.16 | 0.32 | 0.64 | 1.28 | 2.56 |
| Baseline | Clean | MelFB | 34.1 | 55.5 | 70.2 | 78.9 | 84.7 |
| | | LNFB | 18.7 | 32.6 | 53.0 | 69.1 | 79.5 |
| | Reverb | MelFB | 13.3 | 16.3 | 21.7 | 31.1 | 36.4 |
| | | LNFB | 14.0 | 17.7 | 22.2 | 30.1 | 34.8 |
| NMF | Clean | MelFB | 16.4 | 25.5 | 38.9 | 56.3 | 67.8 |
| | | LNFB | 14.3 | 20.8 | 30.6 | 49.6 | 62.5 |
| | Reverb | MelFB | 11.9 | 14.3 | 17.6 | 26.2 | 31.9 |
| | | LNFB | 12.6 | 15.1 | 17.9 | 26.0 | 32.0 |
| SSF | Clean | MelFB | 14.9 | 22.0 | 34.5 | 53.7 | 65.7 |
| | | LNFB | 12.3 | 18.0 | 27.2 | 46.2 | 59.8 |
| | Reverb | MelFB | 11.0 | 12.6 | 15.0 | 21.9 | 26.2 |
| | | LNFB | 11.5 | 12.6 | 15.2 | 21.2 | 25.0 |
| WPE | Clean | MelFB | 9.8 | 19.1 | 39.9 | 61.1 | 72.8 |
| | | LNFB | **7.9** | 13.9 | 29.0 | 53.2 | 67.3 |
| | Reverb | MelFB | 8.7 | **10.0** | **13.1** | **20.0** | **25.5** |
| | | LNFB | 9.8 | 11.4 | 14.2 | 21.0 | 26.3 |

### 3.2. Effect of enhancement techniques

As discussed above, the NMF, SSF and WPE techniques were designed to reduce the mismatch between training and testing conditions. As seen in Table 1, the application of this techniques is always helpful no matter which training data are used. Additionally, we observe that SSF always outperforms NMF for the conditions that we examined. On the other hand, WPE surpasses SSF in all distances only with Reverb training.

The use of WPE in combination with MelFB and Reverb training, and averaging across all RTs, produces the best system for speaker-microphone distances greater than 0.32 m. For the speaker-microphone distance of 0.16 m, the best result is obtained with WPE with Clean training and using the LNFB features. The use of WPE in combination with MelFB and LNFB provides the best results for almost all test conditions, except for the greatest RTs at the longest distances, *i.e.* RT equal to 1.27s and 1.77s at a speaker-microphone distance equal to 2.56 m, where SSF combined with LNFB and Reverb training leads to greater accuracies (see Fig. 1).

### 3.3. Performance of MelFB versus LNFB features

Figure 1 compare directly the best systems obtained using the MelFB and LNFB features. MelFB achieve the best WER in several cases. Nevertheless, as can be seen in Fig. 1, LNFB exhibits better accuracy in some critical RTs and distances, *i.e.* with RT equal to 1.27s and 1.77s at a distance of 2.56 m. On the other hand, LNFB worked better in the shortest distance, *i.e.* 0.16m, for RT equal to 0.84s and 1.27s.

### 3.4. Complementarity between ASR systems

Despite the fact that on average the use of MelFB in combination with WPE and Reverb training provided the lowest WER, different combinations of features, training data and enhancement techniques could address more effectively some testing conditions. The results shown in the Fig. 1 suggest that there is some degree of complementarity between systems trained with different data, enhancement, and parametrization methods. Although we can always select the best system, we can also combine the best engines to obtain a new system that could be even more accurate in different test conditions.

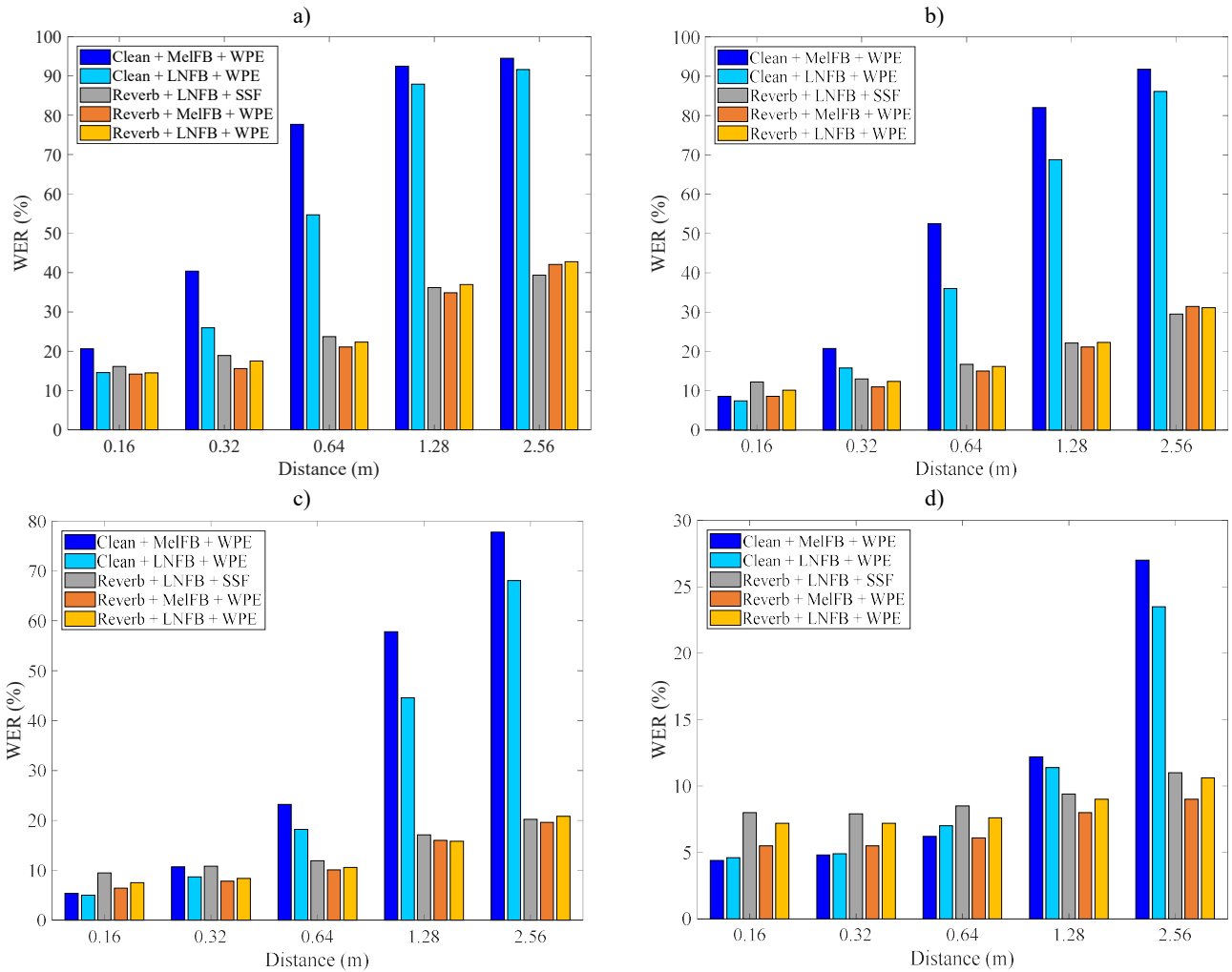

Figure 1: *Results for the best ASR systems with a) RT=1.77 s, b) RT=1.27 s, c) RT=0.84 s and d) RT=0.47s.*

## 4. Conclusions

Two training conditions were evaluated: Clean and Reverb. The comparisons also included the NMF, SSF, and WPE environmental compensation algorithms. The results presented here show that the lowest average WER is achieved using Reverb training and MelFB features combined with WPE. With Clean training, i.e. significant mismatch between testing-training conditions, LNFB features clearly outperform MelFB parameters.

Generally, the use of the NMF, SSF and WPE compensation techniques improves WER for LNFB and MelFB features, for both training styles. Specifically, with Reverb training the use of WPE and LNFB provides WERs that are 3% and 20% lower in average than SSF and NMF, respectively. WPE and MelFB provides WERs that are 11% and 24% lower in average than SSF and NMF, respectively.

It is worth highlighting that for some test conditions some systems led to higher accuracies than MelFB/WP. These results strongly suggest that there is complementarity among the different engines tested here, so finding the best way to combine them is proposed for future research.

## 5. Acknowledgements

The research reported here was funded by Grants Conicyt-Fondecyt 1151306 and ONRG N°62909-17-1-2002. José Novoa was supported by Grant CONICYT-PCHA/Doctorado Nacional/2014-21140711.